\documentclass[preprint,letterpaper,superscriptaddress]{revtex4-1}

\usepackage{amsmath}
\usepackage{amsfonts}
\usepackage{amsbsy}
\usepackage{amssymb}
\usepackage{epsfig}
\usepackage{graphicx}
\usepackage{hyperref}

\newcommand{\abs}[1]{\ensuremath{\left|#1\right|}}
\newcommand{\eref}[1]{(\ref{#1})}

\newcommand{\ud}[0]{\ensuremath{{\rm d}}}

\begin{document}

\title{Metallic Coaxial Nanolasers}

\author{William E. Hayenga}
\author{Hipolito Garcia-Gracia}
\author{Hossein Hodaei}
\affiliation{\footnotesize{CREOL, The College of Optics and Photonics, University of Central Florida, Orlando, FL 32816-2700, USA}}
\author{Yeshaiahu Fainman}
\affiliation{\footnotesize{Department of Electrical and Computer Engineering, University of California, San Diego, California 92093, USA}}
\author{Mercedeh Khajavikhan}
\affiliation{\footnotesize{CREOL, The College of Optics and Photonics, University of Central Florida, Orlando, FL 32816-2700, USA}}
\email[Corresponding author, e-mail: ]{mercedeh@creol.ucf.edu}

\begin{abstract}
The last two decades have witnessed tremendous advancements in the area of nanophotonics and plasmonics. Undoubtedly, the introduction of metallic structures has opened a path towards light confinement and manipulation at the subwavelength scale -- a regime that was previously thought to be out of reach in optics. Of central importance is to devise efficient light sources to power up the future nanoscale optical circuits. Coaxial resonators can provide a platform to implement such subwavelength sources. They support ultrasmall cavity modes and offer large mode-emitter overlap as well as multifold scalability. Given their large modulation bandwidth, they hold promise for high speed optical interconnects -- where they can be used for light generation and modulation simultaneously. In addition, the possibility of thresholdless operation in such devices may have implications in developing the next generation of efficient lighting systems. In this review article, the physics and applications of coaxial nanolasers will be discussed.
\end{abstract}



\maketitle

\section{Introduction}

High-speed data processing is an important aspect of modern society. In recent years, there has been a host of new network configurations to address the rapidly growing need for big data processing and routing. For instance, in order to sustain Moore’s law in electronics, computational systems now employ more parallelism; making communication in such platforms pivotally important \cite{Vishwanath2007}. Another example is the rise of large-scale data centers to power the Internet infrastructure. In order to meet the cost and energy scaling requirements, current data centers must be replaced by an integrated single chip that is networked both internally and externally. This again calls for elaborate and efficient interconnects. Numerous studies, have shown that standard on-chip electrical wires are not well suited to transport such high data-rates, mainly due to their limitations in terms of power consumption and cross-talk \cite{Fainman2013,Miller2000}.

Silicon nanowire waveguides can provide a viable channel for large volumes of data to move at rapid speeds between computer chips in servers. Using complementary metal oxide semiconductor (CMOS)-compatible manufacturing processes, highly integrated optoelectronic circuits can be realized at low cost while meeting the other needs of scalability, bandwidth, fault tolerance, and energy efficiency. In such interconnect schemes, nanolasers, implemented on silicon through heterogeneous integration with III-V semiconductors, can be utilized as sources and perhaps even as modulators \cite{Miller2009}. This has fueled the current interest towards developing the ``ultimate'' nanolaser: a scalable, low threshold source of radiation operating at room temperature, which occupies a small volume on a chip, that can be directly modulated at very high speeds (hundreds of GHz), and is pumped by electric current injection \cite{Noda2006}.

In this manuscript, we summarize some of the advancements in the area of metal-based coaxial nanolasers. First the geometrical structure of the resonator is described, followed by an electromagnetic analysis of its modal content- showing that it is possible to design a cavity with only one subwavelength mode coinciding with the gain bandwidth of the active material. A rate equation model is then provided to explain the threshold behavior as well as the frequency response of such nanolasers. Finally, we describe the fabrication steps involved in implementing these devices and present characterization results. This review will be then concluded with a discussion on future directions.

\vspace{-.5cm}
\subsection{Metal-based nanolasers}

Historically, most efforts towards building ultra-small light sources were focused on dielectric (semiconductor) configurations. In this regard, various constructs have been explored, including vertical Bragg mirror surface emitting cavities, microdisks, and photonic crystals- resulting in some well utilized laser sources such as VCSELs \cite{Iga2000,McCall1992,Painter1999,Park2004,Deppe1997}. In such all-dielectric resonators, the minimum attainable cavity size is fundamentally restricted by diffraction \cite{Coccioli1998}. As a result, the aforementioned arrangements can at best achieve sub-wavelength miniaturization in one or two dimensions.

Due to the plasmonic properties of metals at optical frequencies, metallic nanocavities can support highly confined subwavelength modes \cite{Maier2007}. The first metallic nanolaser demonstrated by Hill et al, was simply comprised of an active nanowire, surrounded by a thin isolating layer, and was coated by gold \cite{Hill2007}. Since then, a plethora of metallic nanolasers have emerged \cite{Mizrahi2008,Hill2009,Nezhad2010,Perahia2009,Yu2010,Lu2010,Ding2011,Ding2011a,Lee2011,Ding2012a,Ding2013a,Kim2011a,Oulton2009,Noginov2009, Kwon2010,Lu2012,Khajavikhan2012,Wu2015}. Most of them, however, support resonance modes that are slightly different from the photonic modes appearing in their non-metallic counterparts \cite{Hill2007,Mizrahi2008,Hill2009,Nezhad2010,Perahia2009,Yu2010,Lu2010,Ding2011,Ding2011a,Lee2011,Ding2012a,Ding2013a,Kim2011a}. This is mainly because such quasi photonic modes minimally interact with the metallic structure -- hence they are negligibly affected by the dissipation losses of metals. Examples of the configurations supporting sub-wavelength photonic modes are metallo-dielectric \cite{Mizrahi2008,Nezhad2010} and nanopatch lasers \cite{Yu2010}. Although being capable of reaching subwavelength dimensions, as they shrink in size, the modes rapidly approach their corresponding cut-off frequencies. This in turn results in a significant drop in the $Q$-factor which eventually makes lasing operation in these configurations altogether unfeasible.

In order to realize deeply subwavelength lasers, the cavity modes must be of the type that could strongly interact with the metal \cite{Oulton2009,Noginov2009,Kwon2010,Lu2012, Khajavikhan2012}. At first glance, the metal loss may appear as a formidable obstacle to reach the lasing threshold in these arrangements. However, the ultra-small mode volumes, unique to this family of resonators, have important ramifications in several aspects of nanolaser design including high mode confinement, large Purcell factor, and in some cases, close to unity spontaneous emission coupling factor. The enhanced atom-field interaction in these resonators can promote lasing action despite the low cavity $Q$-factor. Using this strategy, our group has previously shown ``thresholdless'' lasing in coaxial structures supporting TEM-like modes \cite{Khajavikhan2012}.

Finally it should be noted that while in most of all the above examples, lasing occurs at the near-infrared spectral band, lasing action closer to the metal surface plasmon polariton resonance in the visible range has also been demonstrated  \cite{Oulton2009,Noginov2009,Lu2012}. It is suggested that in some of these light emitting devices, surface plasmons undergo stimulated emission- a claim that is yet to be fully investigated  \cite{Bergman2003}.

\section{Metallic coaxial nanolasers}

The coaxial cavity is a well utilized structure in the microwave regime. In essence, it is a short section of a coaxial waveguide terminated in air. The characteristic impedance mismatch at the interfaces creates virtual mirrors, hence forming a resonator. The coaxial geometry supports the transverse electromagnetic (TEM) mode that has no cut-off frequency - therefore, its cross-section can be scaled down indefinitely while still sustaining a confined mode. It should be noted that in the radio frequency domain this same mode allows energy transport in deeply subwavelength coaxial cables \cite{Cheng1989}. To form a resonator, the waveguide must then be terminated such that its minimum length is approximately $\lambda/(2n_{\rm eff})$ in the longitudinal direction. In the microwave regime, where the perfect conductor approximation for metals holds, the effective index of a coaxial waveguide is identical to the refractive index of its constituent dielectric ring. This makes any further reduction of the resonator length in the axial dimension fundamentally impossible. However, at optical frequencies, due to the plasmonic properties of metals, the effective index of the TEM mode \emph{increases} as the cross-section is reduced -- a trait that allows the coaxial cavity to be scaled down in all three dimensions simultaneously. This unique scalability behavior paves the way towards developing a new class of efficient and high-speed nanoscale light sources based on coaxial cavities \cite{Khajavikhan2012,Feigenbaum2008}.

\begin{figure}[htb]
  \begin{center}
    \includegraphics[width=8.5cm]{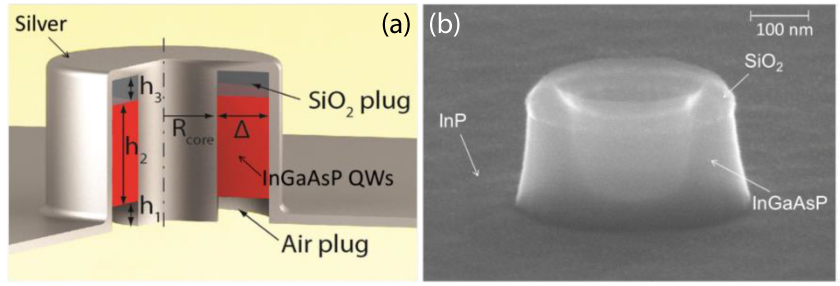}
    \caption{Nanoscale coaxial laser cavity. (a) Diagram of a coaxial laser cavity, the gain medium is shown in red. (b) Scanning electron microscope image of the constituent ring.\label{fig:CoaxialNanolaser}}
  \end{center}
\end{figure}

A coaxial laser resonator is shown in Fig.  \ref{fig:CoaxialNanolaser}a. At the heart of the cavity is a coaxial waveguide that supports plasmonic modes and is composed of a metallic rod enclosed by a metal-coated semiconductor ring. This nanolaser uses a thin dielectric silicon dioxide (SiO$_2$) and an air plug at the top and bottom ends, respectively, as well as an additional metal capping, to improve the mode confinement in this coaxial resonator. The role of the SiO$_2$ plug is to prevent the formation of undesirable plasmonic modes at the top interface, between the metal and the gain medium. The lower air plug is used to allow pump energy into the cavity and also to couple out the generated radiation. The metal is placed in direct contact with the semiconductor to provide a large overlap between the electromagnetic field and the emitters distributed in the volume of the gain medium. In addition, the metallic coating serves as a heat sink that facilitates room temperature and continuous-wave operation.

The above coaxial cavity exhibits several promising properties crucial to the realization of the ``ideal'' nanolaser. First, this structure can support highly confined modes despite being subwavelength -- resulting in a high Purcell factor \cite{Purcell1946} ($F$) as well as a large spontaneous emission coupling into the lasing mode \cite{Yamamoto1991} ($\beta$).  These features in turn can be judiciously incorporated in the design of a nanolaser to reduce the threshold and increase the relaxation oscillation frequency -- potentially eliminating the need for additional stand-alone on-chip modulators. Finally, the reduced radiative lifetime, due to the higher Purcell factor, can lower the total population of carriers that would have otherwise recombined through non-radiative channels. As a result, the coaxial nanolaser is expected to show a lower sensitivity to non-radiative recombination processes that are intrinsic to the semiconductor and its interfaces. These aspects will be further discussed in section  \ref{sec:RateEquation} through rate equation formulations.

\vspace{-.5cm}
\subsection{Modal analysis of coaxial resonators}

The presence of several competing modes in the gain bandwidth of a semiconductor is generally considered undesirable. Below threshold, this will distribute the spontaneous emission power that initiates the lasing process into several channels. Above threshold and especially at higher pump levels, the laser becomes multi-moded -- a process that can cause power fluctuations and instabilities. One of the advantages of reducing the resonator size is that the entailing modal spectrum becomes sparse. It is, therefore, easier to achieve single mode operation in subwavelength cavities.

\begin{figure}[htb]
  \begin{center}
    \includegraphics[width=.85\textwidth]{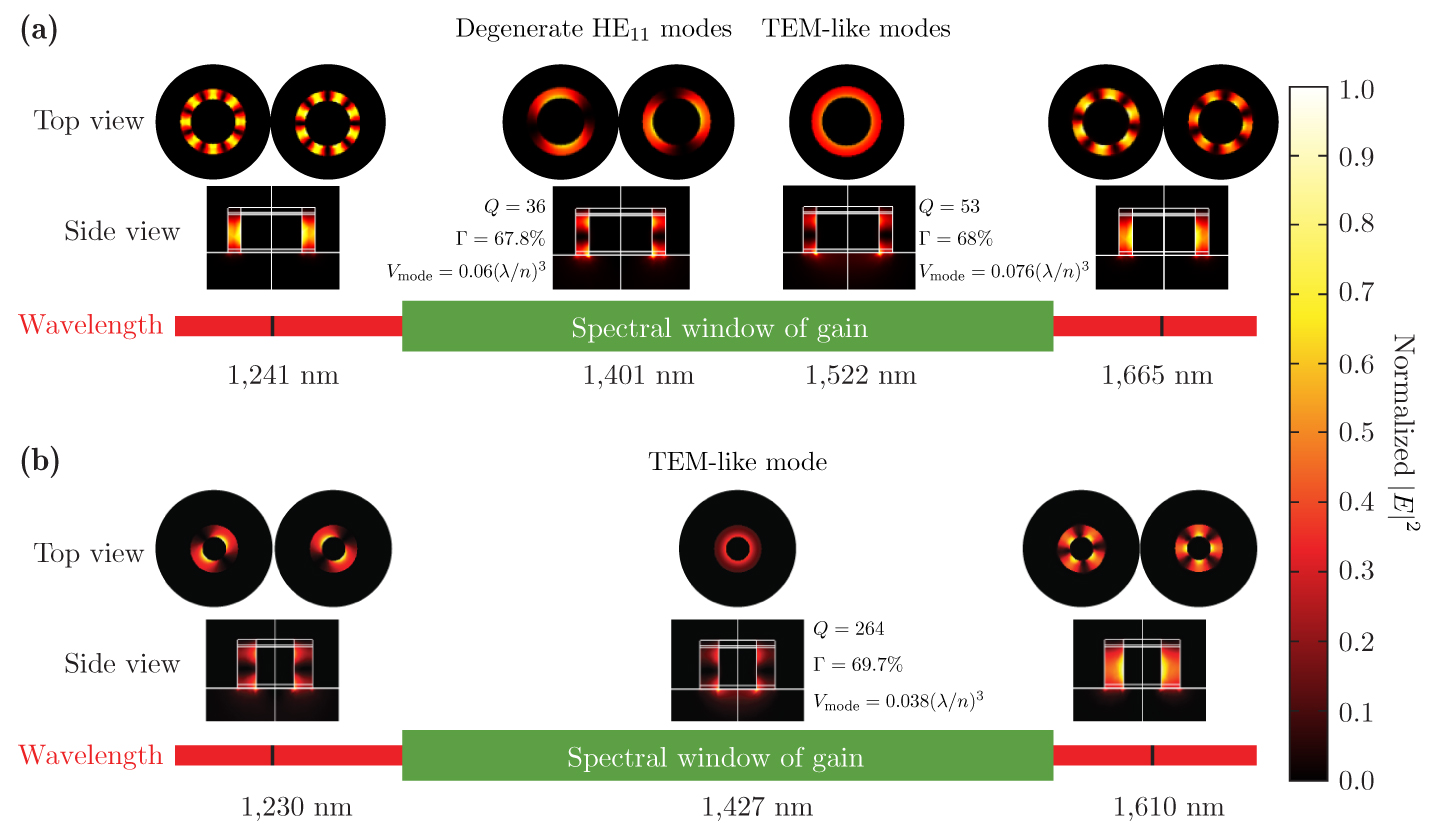}
    \caption{Simulation of the electromagnetic properties of nanoscale coaxial cavities. Supported resonator modes for coaxial nanolasers with inner core radius and gain-medium width of: (a) $R_{\rm core} = 175\textrm{ nm}$ and $\Delta = 75\textrm{ nm}$ at room temperature, and (b) $R_{\rm core} = 100\textrm{ nm}$ and $\Delta = 100\textrm{ nm}$ at 4.5 K, respectively. We use a semiconductor gain medium composed of six quantum wells, resulting in a bandwidth that spans frequencies corresponding to wavelengths in vacuum from 1.26 to 1.59 $\mu$m at room temperature, and from 1.27 to 1.53 $\mu$m at 4.5 K.\label{fig:modes}}
  \end{center}
\end{figure}

Figure \ref{fig:modes} shows how small changes in cavity size can affect the modal content of nanoscale coaxial lasers. In this figure, two different configurations are depicted: one has an inner core cylinder (radius ($R_{\rm core}$): 175 nm), a gain-medium (ring thickness ($\Delta$): 75 nm, height ($h_{2}$): 210 nm), and a lower and an upper plug with heights of $h_{1} = 20\textrm{ nm}$ and $h_{3} = 30\textrm{ nm}$, respectively. The second structure shares the same plug and gain medium heights, but it is smaller in diameter, with $R_{\rm core} = 100\textrm{ nm}$ and $\Delta = 100\textrm{ nm}$. In order to determine their modal spectra, electromagnetic simulations based on finite element method (FEM) are carried out. Figure \ref{fig:modes}a shows that for the first arrangement the fundamental TEM-like mode as well as the two degenerate $\textrm{HE}_{11}$ modes fall within the gain bandwidth of the active medium (in this case the gain material is comprised of six InGaAsP quantum wells). The second structure, on the other hand, only supports the fundamental TEM-like mode that has a higher quality factor than any of the supported modes in the first design \cite{Khajavikhan2012}.

The Purcell factor ($F$) represents the enhancement of the radiative decay rate of an emitter confined within a cavity environment to that of the bulk,
%
\begin{equation}\label{eq:PurcellFactor}
    F = \frac{3}{4\pi^{2}}\left(\frac{\lambda_{0}}{n}\right)^{3}\left(\frac{\min\{Q,Q_{\rm h}\}}{V_{\rm m}}\right),
\end{equation}
where $(\lambda_{0}/n)$ is the wavelength in the material, $Q_{\rm h}$ is the quality factor associated with the homogeneous lineshape of the gain medium, and $Q$ and $V_{\rm m}$ are the quality factor and mode volume of the cavity, respectively. Similarly, the spontaneous emission coupling factor $\beta$ is defined as the ratio of spontaneous emission into the lasing mode to the total spontaneous emission generated by the gain medium, and can be written as
\begin{equation}\label{eq:beta}
  \beta = \frac{F^{(1)}}{\sum_{k}F^{(k)}},
\end{equation}
where the lasing mode is indicated by $k=1$, and the summation is over all modes, including confined cavity modes as well as the free space radiation modes. The spontaneous emission coupling factor can be brought close to the theoretical limit of unity by increasing the Purcell factor associated with the lasing mode, reducing the number of cavity modes that coincide with the gain spectrum, and finally by blocking the coupling to the radiation modes through the use of a metallic cladding. $\beta$ can be calculated by placing a number of randomly distributed dipoles in the active region and by computing their emitted power at different wavelengths. The spontaneous emission coupling factor can then be estimated by the ratio of the emitted power that spectrally coincides with the lasing mode to the total power radiated by the dipoles.  Following this procedure, the single-mode cavity of Fig. \ref{fig:modes}b exhibits a near unity spontaneous emission coupling factor ($\beta \approx 0.99$).  This condition is usually referred to as ``thresholdless'' lasing.

\vspace{-.5cm}
\subsection{Rate-equation model\label{sec:RateEquation}}

At first glance, the low $Q$-factor of the resonator may imply that a nanoscale coaxial cavity requires a high pump power to reach the onset of lasing. However, a careful analysis shows that due to the high mode confinement ($\Gamma$), close to unity spontaneous emission coupling ratio, and large Purcell factor ($F$), such structures can indeed display both low threshold and unprecedentedly broad modulation bandwidths -- two traits that are usually at odds with one another. Here we introduce a simplified rate-equation model that could further elucidate the laser dynamics in such arrangements:
\begin{subequations}\label{eq:rateEqs}
  \begin{align}
    \frac{\ud n_{\rm c}}{\ud t} &= \frac{I}{q}- \frac{F\beta}{n_{\rm sp}\tau_{\rm sp}}\Gamma n_{\rm p} n_{\rm c} - \frac{F}{\tau_{sp}}n_{\rm c} - \frac{1}{\tau_{\rm nr}}n_{\rm c}, \label{eq:rateEq1}\\
    \frac{\ud n_{\rm p}}{\ud t} &=  \frac{F\beta}{\tau_{\rm sp}n_{\rm sp}}\Gamma n_{\rm p}n_{\rm c} + \frac{F}{\tau_{\rm sp}}\beta n_{\rm c} - \frac{1}{\tau_{\rm p}}n_{\rm p}. \label{eq:rateEq2}
  \end{align}
\end{subequations}

 Here $n_{\rm c}$ and $n_{\rm p}$ are the total electron-hole pair and the number of photons, $I$ is the injection current, $q$ is the electron charge, and $\tau_{\rm sp}$ and $\tau_{\rm nr}$ are the bulk spontaneous and non-radiative lifetimes. The cavity lifetime is $\tau_{\rm p} = Q\lambda/2\pi c$, where $\lambda$ is the wavelength and $c$ the speed of light in vacuum, Lastly $n_{\rm sp}=(f_{\rm c} (1-f_{\rm v}))/(f_{\rm c}-f_{\rm v})$, where $f_{\rm c}$ and $f_{\rm v}$ are the Fermi-Dirac functions in the conduction and valence bands \cite{Yamamoto1991,Ni2012a,Khurgin2014}.

At steady-state, the threshold current and the number of photons in the laser mode are given by:
\vspace{-.25cm}
\begin{subequations}
\begin{align}
  I_{\rm th} &= \frac{2\pi c n_{\rm sp}}{\lambda} \frac{q}{\beta\Gamma Q} \left(1 + \frac{\tau_{\rm sp}}{F \tau_{\rm nr}}\right),\label{eq:Ith}\\
  n_{\rm p} &= \frac{1}{2}\left[ \frac{I}{q}\tau_{\rm p}-\frac{n_{\rm sp}}{\beta\Gamma} \left(1+\frac{\tau_{\rm sp}}{F\tau_{\rm nr}}\right)\right] + \sqrt{\frac{I}{q}\tau_{\rm p} \frac{n_{\rm sp}}{\Gamma} + \frac{1}{4}\left[\frac{I}{q}\tau_{\rm p} - \frac{n_{\rm sp}}{\beta\Gamma} \left(1+\frac{\tau_{\rm sp}}{F\tau_{\rm nr}}\right)\right]^{2}}.\label{eq:photonNumber}
\end{align}
\end{subequations}

It can be seen from Eq. \eref{eq:Ith} that the threshold current is inversely proportional to the product of $\beta$, $\Gamma$ and $Q$. On the other hand, the Purcell factor contributes in reducing the threshold only through increasing the decay rate of the spontaneous emission in respect to that of the non-radiative processes. However, in most semiconductor gain systems where $\tau_{\rm sp}/\tau_{\rm nr} \gg 1$, a large Purcell factor is expected to play a significant role in determining the threshold value.

\begin{figure}
    \begin{center}
    \includegraphics[width=.8\textwidth]{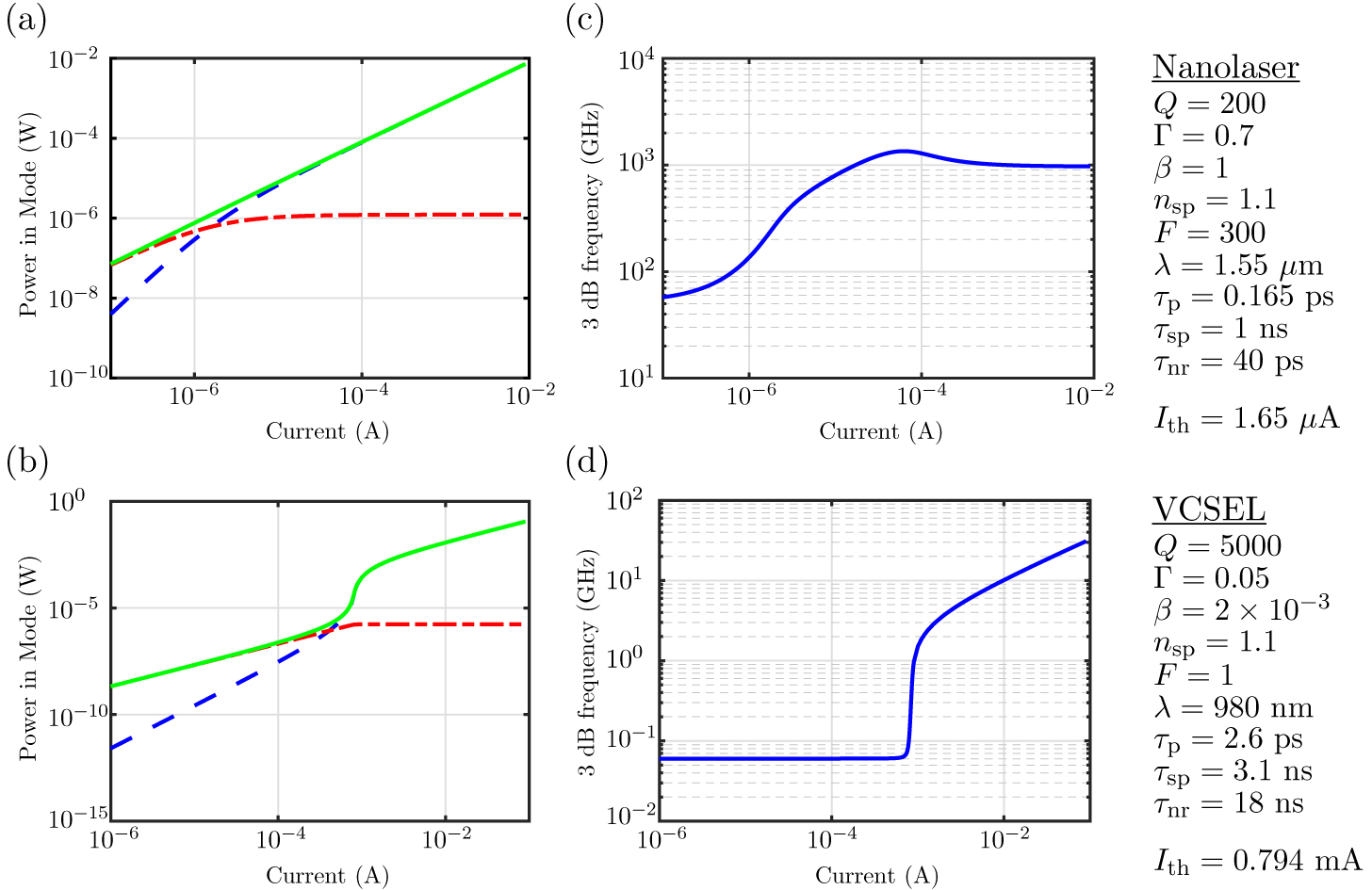}
    \caption{Light-Current curve for a thresholdless coaxial nanolaser (a) and a VCSEL (b). Green (solid)is the total power in the mode, blue (dashed) the stimulated power contribution, and red (dash-dotted) the spontaneous power contribution. Using Eq. \eref{eq:Ith}, the threshold current of the nanolaser and the VCSEL are 1.65 $\mu$A and 0.794 mA, respectively. Plot of the 3dB modulation frequency as a function of injection current for the nanolaser (c) and VCSEL (d). The nanolaser exhibits a modulation response of over 1 THz.\label{fig:LvsI_f3dB}}
    \end{center}
\end{figure}

A light-emitting device is considered to be a laser if stimulated emission dominates the output power. Typically, for a laser, the Light-Injection (L-I) or Light-Light (L-L) curve in the logarithmic-logarithmic (log-log) scale is expected to consist of three distinct regions. At low pump intensities, the device is in the photoluminescence (PL) regime, where it operates as an LED and the slope of the L-I or L-L curve is unity. At higher pump levels, it enjoys a sudden increase in the power allocated to the future lasing mode -- creating a nonlinear kink that is known as the amplified spontaneous emission regime (ASE) and the slope of the log-log curve is larger than one. By further increasing the pump energy, the slope eventually returns to unity and the device operates in the lasing mode. Figures \ref{fig:LvsI_f3dB}a,b display the Light-Current (L-I) curves for two lasers: one is the coaxial nanolaser as presented in Fig. \ref{fig:CoaxialNanolaser}b with a spontaneous emission coupling factor of unity ($\beta\sim1$), and the other one is a typical microscale laser (parameters are extracted from \cite{Lorke2013} for a vertical cavity surface emitting laser (VCSEL)). While in simulation, the trajectory of the spontaneous and stimulated emission curves (dash-dotted and dashed lines) can specify the threshold power, for a device with $\beta\rightarrow1$, the lack of a discernible kink in the total output power makes it virtually impossible to determine the threshold via L-I or L-L measurements. Such a soft transition from PL to lasing is known as ``thresholdless'' behavior. Using the parameters given in Fig. \ref{fig:LvsI_f3dB}, the corresponding lasing thresholds for the coaxial nanolaser and the VCSEL are 1.65 $\mu$A ($\sim$2 kA/cm$^2$) and 0.794 mA ($\sim$700 A/cm$^{2}$), respectively.

The direct modulation behavior of a single mode laser diode can be derived from a small signal analysis ($n_{\rm c}\rightarrow n_{\rm c}+\Delta n_{\rm c}$, $n_{\rm p}\rightarrow n_{\rm p}+\Delta n_{\rm p}$, and $I \rightarrow I+\Delta I$) performed on Eq. \eref{eq:rateEqs} \cite{Coldren1995}. The frequency of a system is defined as

\begin{equation}\label{eq:modResp}
    H(f) = \frac{f_{\rm r}^2}{f_{\rm r}^2 - f^2 + i\zeta f}.
\end{equation}
Here $f_{\rm r}$ is the relaxation frequency and $\zeta$ is the damping coefficient, which are given as below
\begin{subequations}
\begin{align}
	(2\pi f_{\rm r})^{2} &= \left(\frac{F\Gamma\beta n_{\rm p}}{\tau_{\rm sp} n_{\rm sp}}+\frac{F\beta}{\tau_{\rm sp}}\right)\left(\frac{F\Gamma\beta n_{\rm c}}{\tau_{\rm sp} n_{\rm sp}}\right) +  \left(\frac{1}{\tau_{\rm p}}-\frac{F\Gamma\beta n_{\rm c}}{\tau_{\rm sp} n_{\rm sp}}\right)\left(\frac{F\Gamma\beta n_{\rm p}}{\tau_{\rm sp} n_{\rm sp}} + \frac{F}{\tau_{\rm sp}} + \frac{1}{\tau_{\rm nr}}\right),\label{eq:fr}\\
	2\pi\zeta &=  \left(\frac{1}{\tau_{\rm p}}-\frac{F\Gamma\beta n_{\rm c}}{\tau_{\rm sp} n_{\rm sp}}\right) + \left(\frac{F\Gamma\beta n_{\rm p}}{\tau_{\rm sp} n_{\rm sp}} + \frac{F}{\tau_{\rm sp}} + \frac{1}{\tau_{\rm nr}}\right).\label{eq:zeta}
\end{align}
\end{subequations}
Of particular importance is to find the 3dB frequency ($f_{\rm 3dB}$), where $\abs{H(f)}^2 = 1/2$,
\begin{equation}\label{eq:f_3dB}
    f_{\rm 3dB} = \sqrt{f_{\rm r}^2-\frac{\zeta^2}{2} + \frac{1}{2} \sqrt{8f_{\rm r}^4-4f_{\rm r}^2 \zeta^2+\zeta^4}}.
\end{equation}

Figures \ref{fig:LvsI_f3dB}c and d show the 3dB bandwidth as a function of the injection current for the nanolaser and the VCSEL. While at ten times the threshold power, a typical microscale VCSEL is expected to show a modulation bandwidth of up to 30 GHz, a coaxial nanolaser can potentially show a much larger bandwidth on the order of a few hundreds of GHz. Such unprecedented large modulation bandwidth is perhaps one of the most compelling drives behind the research towards developing such nanoscale lasers.

\vspace{-.25cm}
\subsection{Fabrication method and characterization}

\begin{figure}[htb]
  \begin{center}
    \includegraphics[width=0.85\textwidth]{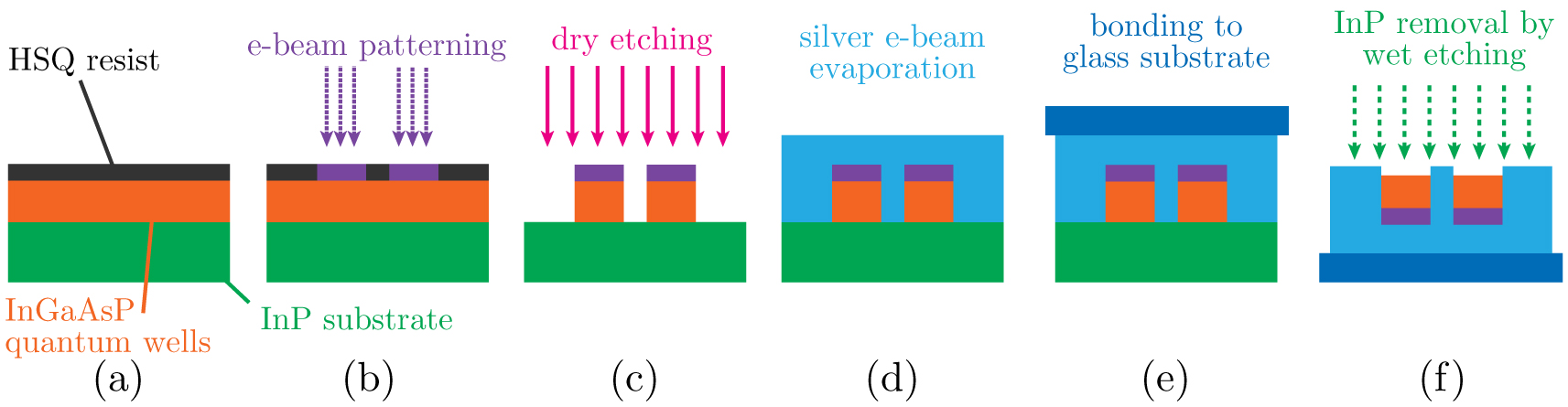}
    \caption{Main fabrication steps involved in implementing a coaxial nanolaser\label{fig:fabrication}}
  \end{center}
\end{figure}

The optically pumped nanoscale coaxial cavities can be fabricated using standard nano-fabrication techniques. The main steps involved in the implementation of these lasers are depicted in Fig. \ref{fig:fabrication}.  The gain medium consists of six quantum wells with an overall height of 200 nm, each composed of a 10 nm thick well (In$_{x=0.56}$Ga$_{1-x}$As$_{y=0.938}$P$_{1-y}$) sandwiched between two 20 nm thick barrier layers (In$_{x=0.734}$Ga$_{1-x}$As$_{y=0.57}$P$_{1-y}$). The quantum wells are grown on a p-type InP substrate and are covered by a 10 nm thick InP overlayer for protection. A hydrogen silsesquioxane (HSQ) solution in methyl isobutyl ketone (MIBK) is used as a negative tone inorganic electron beam resist (Fig. \ref{fig:fabrication}a). Rings with different inner radii and widths are written by electron beam exposure (Fig. \ref{fig:fabrication}b) -- where the exposed HSQ serves as a mask for a subsequent reactive ion etching process. After dry etching (Fig. \ref{fig:fabrication}c), the samples are studied under a scanning electron microscope (Fig. \ref{fig:CoaxialNanolaser}b). Subsequently, a silver coating is deposited using e-beam evaporation technique (Fig. \ref{fig:fabrication}d). The sample is then glued, upside down, to a piece of glass (Fig. \ref{fig:fabrication}e) with silver epoxy and dipped in hydrochloric acid in order to remove the InP substrate and to open up the air-plug aperture (Fig. \ref{fig:fabrication}f) \cite{Khajavikhan2012}.

\vspace{-.25cm}

The coaxial nanolasers are characterized under optical pumping ($\lambda_{\rm pump} = 1064\textrm{ nm}$) in continuous wave (CW) and pulsed regime. A microscope objective is used to focus the pump beam, as well as to collect the output light from the samples. Output spectra are obtained using a monochromator in conjunction with a cooled InGaAs detector. If necessary, continuous wave and pulsed mode measurements were performed with adequate overlap to ensure that the two sets of data were consistent after scaling \cite{Khajavikhan2012}.

\begin{figure}[htb]
  \begin{center}
    \includegraphics[width=.95\textwidth]{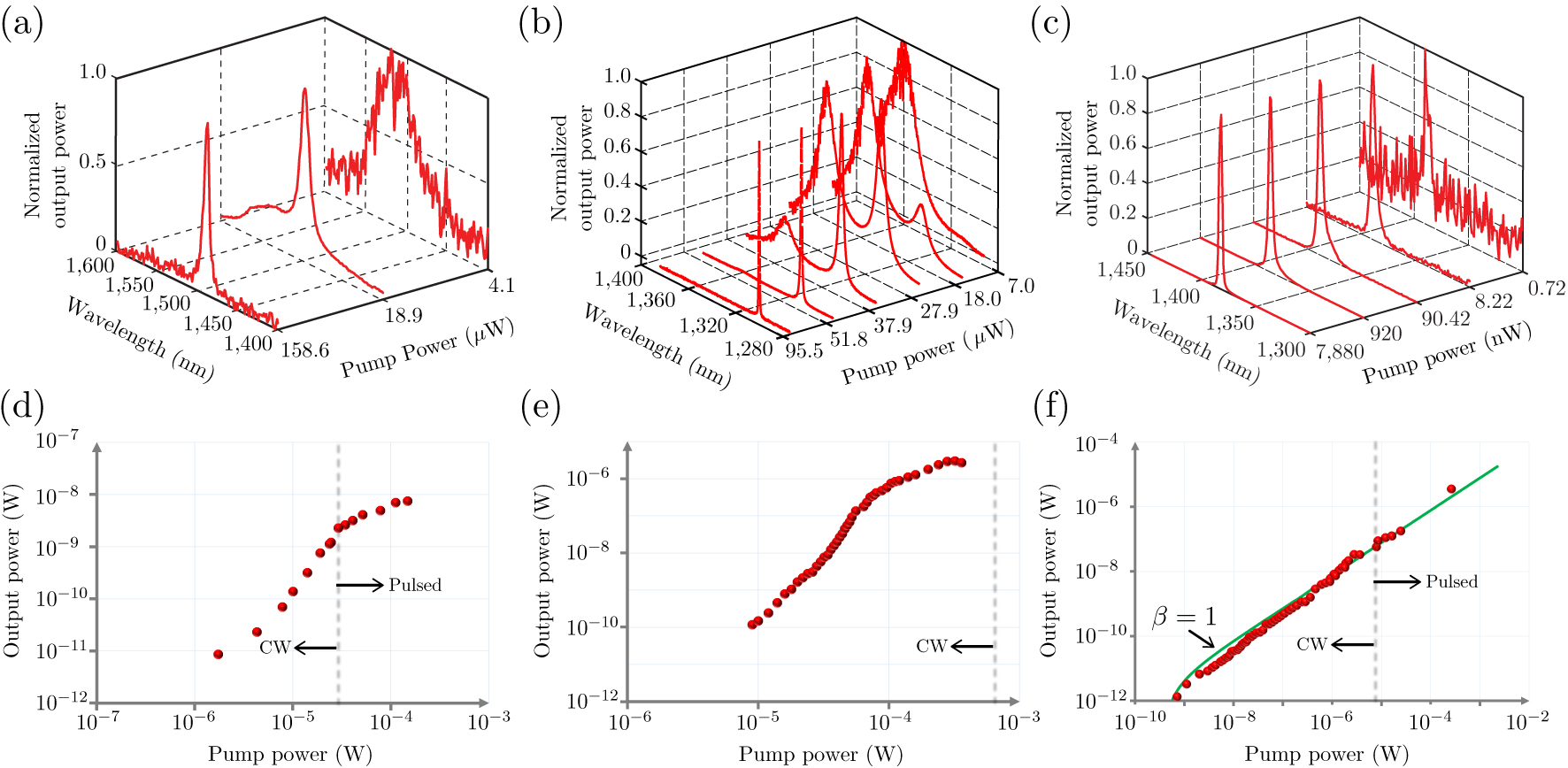}
    \caption{Emission characteristics for three nanoscale lasers operating at (a,d) room temperature ($R_{\rm core}$: 175 nm, $\Delta$:75 nm), (b,e) 77 K ($R_{\rm core}$: 50 nm, $\Delta$: 200 nm), and (c,f) 4.5 K ($R_{\rm core}$: 100 nm, $\Delta$:100 nm). Shown are spectral evolution (a-c) and the light-light curve (d-f).\label{fig:characterization}}
  \end{center}
\end{figure}

Figure \ref{fig:characterization} shows the emission characteristics of three different nanoscale coaxial lasers operating at room temperature ($R_{\rm core}$: 175 nm, $\Delta$:75 nm), 77 K ($R_{\rm core}$: 50 nm, $\Delta$: 200 nm), and 4.5 K ($R_{\rm core}$: 100 nm, $\Delta$:100 nm). Figures \ref{fig:characterization}a, b and c display the normalized evolution spectra of these three devices, respectively. Of interest in these plots is the emission linewidth that narrows down considerably as the pump power increases.  A closer look at the Light-Light curves reveals that the two lasers in Figs. \ref{fig:characterization}d and e show standard laser behavior, where spontaneous emission dominates at lower pump powers (photoluminescence region) and stimulated emission is dominant at higher pump powers (lasing region). The photoluminescence and lasing regions are connected through a pronounced amplified spontaneous emission (ASE) transient region. The situation however changes drastically for the structure reported in Figs. \ref{fig:characterization}c and f, as the light-light curve shown in Fig. \ref{fig:characterization}f follows a straight line with no pronounced kink. This agrees well with the thresholdless lasing operation predicted from the electromagnetic analysis in Fig. \ref{fig:modes}b, where only a single non-degenerate cavity mode exists within the bandwidth of the gain medium. The thresholdless behavior is also evident from the spectral evolution, seen in Fig. \ref{fig:characterization}c, where a single narrow Lorentzian-like emission peak is obtained throughout the whole range of pump powers. A second order coherence $g^{(2)}$ measurement recently performed on the above devices further confirms the lasing behavior in such nanoscale coaxial structures \cite{HayengaCLEO2016}.

\vspace{-.75cm}
\section{Discussion and future work}

In this paper, recent advances in the area of nanoscale metallic coaxial lasers are reviewed. The applications of nanolasers range from on-chip optical communication to high-resolution and high-throughput imaging, sensing and spectroscopy. This has fueled interest in developing the ``ultimate'' nanolaser: a scalable, low-threshold source of radiation that operates at room temperature and occupies a small volume on a chip. Coaxial nanoresonators can provide a platform to implement such subwavelength sources. They support ultra-confined cavity modes and offer large mode-emitter overlap as well as multifold scalability. In this manuscript, the modulation response of these lasers is further discussed. The rate equation model developed in Section  \ref{sec:RateEquation} shows that coaxial nanolasers can be modulated at speeds in excess of a few hundred gigahertz. In addition, thresholdless operation can facilitate light generation in these lasers. The possibility of thresholdless lasing may have some implications in developing the next generation of efficient lighting systems \cite{Chow2015}.

So far the demonstration of coaxial nanolasers has been limited to optically pumped devices. In order to enable electrical pumping, the semiconductor layer structure must be modified to incorporate appropriate p-n junctions. In addition, a mechanism for electrically isolating the device from the inner and outer metal claddings has to be considered. Another important issue is the management of heat generated by current injection. As electrically pumped devices, this family of nanolasers is expected to play an important role in the future photonic integrated circuits.

\vspace{-.65cm}
\section*{Acknowledgements}

The authors greatly appreciate the support from Army Research Office (grants: W911NF-16-1-0013 and W911NF-14-1-0543). The authors would also like to thank the support from the Defense Advanced Research Projects Agency (DARPA), the National Science Foundation (NSF), the NSF Center for Integrated Access Networks (CIAN), and the Cymer Corporation.


\end{document}